\documentclass[twocolumn]{openjournal}
\usepackage{graphicx} 
\usepackage{multirow}

\usepackage[colorlinks,linkcolor=blue,citecolor=blue,urlcolor=blue ]{hyperref}
\usepackage[utf8]{inputenc}
\usepackage{float}
\usepackage{xcolor}
\usepackage{ulem}
\usepackage[T1]{fontenc}

\def\sigh2{$\Sigma_{\rm H_2}$}

\def\dsfr{$\Delta$SFR}
\def\kms{km~s$^{-1}$}

\begin{document}
\title{Galaxy evolution in the post-merger regime.  II - Post-merger quenching peaks within 500 Myr of coalescence\vspace{-1.5cm}}
\author{Sara L. Ellison$^1$}
\author{Leonardo Ferreira$^1$}
\author{Vivienne Wild$^2$}
\author{Scott Wilkinson$^1$}
\author{Kate Rowlands$^3$}
\author{David R. Patton$^4$}
\thanks{$^*$E-mail: sarae@uvic.ca}
\affiliation{$^1$ Department of Physics \& Astronomy, University of Victoria, Finnerty Road, Victoria, BC V8P 1A1, Canada\\
  $^2$ School of Physics and Astronomy, University of St Andrews, North Haugh, St Andrews, KY16 9SS, U.K.\\
  $^3$  AURA for ESA, Space Telescope Science Institute, 3700 San Martin Drive, Baltimore, MD 21218, USA\\
  $^4$ Department of Physics and Astronomy, Trent University, 1600 West Bank Drive, Peterborough, ON K9L 0G2, Canada
}

\begin{abstract}
  Mechanisms for quenching star formation in galaxies remain hotly debated, with galaxy mergers an oft-proposed pathway. In Ellison et al. (2022) we tested this scenario by quantifying the fraction of recently and rapidly quenched post-starbursts (PSBs) in a sample of post-merger galaxies identified in the Ultraviolet Near Infrared Optical Northern Survey (UNIONS).  Compared with a control sample of non-interacting galaxies, Ellison et al. (2022) found PSBs to be a factor of 30-60 more common in the post-mergers, demonstrating that mergers can lead to quenching.  However, the exact timing of this post-merger quenching was unconstrained.  Thanks to our recent development of the Multi-Model Merger Identifier (\textsc{mummi}) neural network ensemble (Ferreira et al. 2024a,b), we are now able to predict the time since coalescence ($T_{PM}$) for the UNIONS post-merger galaxies up to $T_{PM} = 1.8$ Gyr, allowing us to further dissect the merger sequence and measure more precisely when quenching occurs.  Based on a sample of 5927 $z<0.3$ post-mergers identified in UNIONS, we find that the post-coalescence population evolves from one dominated by star-forming (and starbursting) galaxies at $0 < T_{PM} < 0.16$ Gyr, through to a population that is dominated by quenched galaxies by $T_{PM} \sim 1.5$ Gyr.  By combining the post-mergers with a sample of 15,831 spectroscopic galaxy pairs with projected separations $r_p<100$ kpc we are able to trace the evolution of quenching during the full merger sequence.  We find a PSB excess throughout the post-merger regime, but with a clear peak at $0.16 < T_{PM} < 0.48$ Gyr.  In this post-merger time range PSBs are more common than in control galaxies by factors of 30-100 (depending on PSB selection method), an excess that drops sharply at longer times since merger.    We also quantify the fraction of PSBs that are mergers and find that the majority (75 per cent) of classically selected E+A are identified as either pairs or post-mergers, with a lower merger fraction (60 per cent) amongst PCA selected PSBs.  The merger fraction of PSB galaxies also correlates strongly with stellar mass.  Taken together, our results demonstrate that 1) galaxy-galaxy interactions can lead to rapid post-merger quenching within 0.5 Gyr of coalescence, 2) the majority of (but not all) PSBs at low $z$ are linked to mergers and 3) quenching pathways are diverse, with different PSB selection techniques likely identifying galaxies quenched by different physical processes with an additional dependence on stellar mass.
  
\end{abstract}
\maketitle

\section{Introduction} \label{intro_sec}

The study of galaxy mergers has a rich observational history rooted in visual characterizations.  Arguably, the godfather of interacting galaxies was Halton Arp, who compiled a catalog of `peculiar' objects into his now famous atlas (Arp et al. 1966).  Although many automated classification systems have since been developed (e.g. Conselice 2003; Lotz et al. 2004; Pawlik et al. 2016; Nevin et al. 2019), the popularity of visual classifications has persisted.  These efforts range from the solitary (and probably exhausted) scrutineer (e.g. Bridge et al. 2010; Nair \& Abraham 2010), to the professional team effort (e.g. Kartaltepe et al. 2015; Verrico et al. 2023) and ultimately to crowd-sourced citizen science (Darg et al. 2010; Casteels et al. 2013; Willett et al. 2013; Tanaka et al. 2023).  Many of these works have attempted not only to identify an interaction, but also to place it along a merger sequence, i.e. distinguishing pre-pericentric cases from those that have already had one close passage and those that have fully coalesced (Toomre 1977; Brassington, Ponman \& Read 2007; Knapen et al. 2014; Kartaltepe et al. 2015; Barrera-Ballesteros et al. 2015; Pan et al. 2019).  

\medskip

\subsection{Galaxy evolution along the merger sequence}

The motivation for assembling an interaction timeline lies in the desire to pinpoint not only the onset, but also the persistence of, merger induced events such as starbursts and nuclear accretion.  However, assembling an accurate merger sequence based purely on imaging is notoriously difficult (Soares 2007; Blumenthal et al. 2020).  Fortunately, in the pre-coalescence regime, studies of galactic orbits can be used to translate inter-galactic separations into timescales until merger (e.g. Brassington et al. 2007; Patton et al. 2024).  Indeed, numerous observational works have studied the star formation rates (e.g. Nikolic, Cullen \& Alexander 2004; Ellison et al. 2008; Li et al. 2008; Patton et al. 2013; Scott \& Kaviraj 2014), active galactic nucleus (AGN) fraction (Ellison et al. 2011, 2013, 2019a; Satyapal et al. 2014; Ellison, Patton \& Hickox 2015; Weston et al. 2017; Goulding et al. 2018; Bickley et al. 2023, 2024a), metallicity (Kewley et al. 2006; Michel-Dansac et al. 2008; Scudder et al. 2012; Bustamante et al. 2020) and morphological transformation (e.g. De Propris et al. 2007; Casteels et al. 2014; Patton et al. 2016) of galaxy pairs, generally finding coherent trends with pair separation. Despite the imperfect connection between separation and time, these observational results nonetheless agree qualitatively with predictions from simulations (e.g. Rupke et al. 2010; Torrey et al. 2012; Patton et al. 2013, 2020).  Additional information on the internal dynamics further helps to fine-tune the pre-coalescence timing (Soares 2007; Hung et al. 2016; Feng et al. 2020).

\smallskip

Whereas the pair regime is extremely well studied in observations, the post-coalescence phase has received relatively little attention.  This is unfortunate, because many of the merger-induced effects seen in the pair phase, such as starbursts and AGN triggering, are predicted to peak at, or soon-after, coalescence (e.g. Torrey et al. 2012; Hani et al. 2020; McAlpine et al. 2020; Byrne-Mamahit et al. 2024).  Moreover, one of the most important predictions from simulations, namely that the merger will eventually lead to the cessation of star formation, might predominantly be seen only \textit{after} coalescence (Hopkins et al. 2008; Snyder et al. 2011; Pawlik et al. 2019).

\subsection{The advent of machine learning}

Historically, the dearth of characterization of post-mergers can be linked back to the challenge of their identification.  Whereas finding galaxy pairs can be done in an automated way using only sky positions and redshifts, finding post-mergers requires a morphological assessment that is either time-consuming by eye, or highly incomplete using traditional automated techniques (e.g. Wilkinson et al. 2024).  Therefore, pure samples of post-mergers are relatively sparse and few studies of the post-coalescence phase have been conducted in a statistical way (e.g. Ellison et al. 2013; Li et al. 2023a).  However, with the advent of machine vision techniques, post-merger galaxy identification has boomed.  Numerous studies have developed deep learning methods to identify mergers in various stages, including post-coalescence, yielding samples of hundreds or even thousands of galaxies (Pearson et al. 2019a, 2022; Ferreira et al. 2020;  Bickley et al. 2021; Ferreira et al., 2022; Walmsley et al. 2022; Omori et el. 2023).  The resulting statistical studies are now giving us the first robust measurements of post-merger AGN and star formation activity (Pearson et al. 2019b; Bickley et al. 2022, 2023).

\smallskip

Although these recent studies of post-merger galaxies have provided us with one more data point on the interaction timeline, these samples are likely to contain a diverse population of post-merger galaxies whose time since coalescence may span a gigayear or more. The holy grail in the assembly of a detailed merger sequence is the assessment of time since coalescence of a given post-merger galaxy.   It is in this task that machine vision has the potential to take us into a truly new era.  Whereas merger \textit{identification} (Pearson et al. 2019a, 2022; Ferreira et al. 2020;  Bickley et al. 2021; Walmsley et al. 2022; Omori et el. 2023) has largely been used to speed up a procedure that can readily done by the human eyeball, quantifying the time-since-merger is not a task that has ever been attempted manually (at least not beyond a few crude and qualitative bins, or time intensive modelling of case studies, e.g. Toomre 1977; Barnes 2004; Privon et al. 2013; Mortazavi et al. 2016; Calderon-Castillo et al. 2024).  However, initial studies have indicated that time-labelling post-mergers with neural networks may be possible (Koppula et al. 2021; Pearson et al. 2024). If perfected, the labelling of galaxies by their time-since-merger will allow a fast, automated and detailed analysis of the persistence of changes induced by the interaction.

\smallskip

In Ferreira et al. (2024a) we introduced a hybrid method combining convolutional neural networks (CNNs) and vision transformers, the Multi-Model Merger Identifier (\textsc{mummi}), that achieves some of the highest purity post-merger classifications to date.  \textsc{mummi} is trained and tested on galaxies drawn from the IllustrisTNG simulation where the merger status is known and hence model performance can be assessed (more details on \textsc{mummi} are provided in Section \ref{mummi_sec}).  One novel component of \textsc{mummi} is the adoption of a jury approach, in which an ensemble of 20 networks produces an independent merger probability that can be combined for higher confidence.  This is similar in spirit to the way that multiple human votes are combined in either citizen science projects, or amongst groups of experts (e.g. Darg et al. 2010; Kartaltepe et al. 2015; Bickley et al. 2021).  In Ferreira et al. (2025b) we extended our methodology to predict the time-since-merger of galaxies identified by \textsc{mummi} and present a sample of $\sim$ 11,000 majority-identified post-mergers in the Ultraviolet Near Infrared Optical Northern Survey (UNIONS) with log $(M_{\star}/M_{\odot}) > 10$.  With this time labelled post-merger catalog in hand, we are now finally able to assess merger-induced effects throughout the interaction sequence.  In a series of papers, we present the time sequence, from pairs to $\sim$ 1.8 Gyr post-coalescence, of star formation rate enhancement (Ferreira et al. 2025a), quenching of star formation (this work) and AGN triggering (Ellison et al. 2025).


\subsection{The merger-quenching connection}

The idea that mergers might lead to quenching has been largely motivated by early simulations with aggressive feedback recipes that followed the triggering of quasar activity in the post-merger regime (Di Matteo, Springel \& Hernquist 2005; Hopkins et al. 2008; Cattaneo et al. 2009).  However, the revised recipes of AGN feedback implemented in modern cosmological simulations have prompted a re-evaluation of the connection between mergers and quenching.  Whereas some simulations continue to find a link between mergers and the cessation of star formation (e.g. Davis et al. 2019; Lotz et al. 2021), others find no connection (Rodriguez-Montero et al. 2019).  In practice, quenching is likely to be a process with multiple pathways (e.g. Schawinski et al. 2014;  Pawlik et al. 2018; 2019) and mergers may only quench star formation when the conditions are `just right' (Snyder et al. 2011; Zheng et al. 2020).  Overall though, quenched post-mergers in simulations are rare, but nonetheless in excess compared with control (non-interacting) galaxies (Quai et al. 2021; 2023).

\smallskip

Observationally, numerous works have attempted to link mergers to quenching via the assessment of galactic gas content.  The motivation here has been that feedback must first disrupt the gas content before it shuts down star formation.  Neither studies of the atomic (Dutta et al. 2018, 2019; Ellison, Catinella \& Cortese 2018; Diaz-Garcia \& Knapen 2020; Bok et al. 2022), molecular (Casasola et al. 2004; Pan et al. 2018; Violino et al. 2018; Lisenfeld et al. 2019; Yu et al. 2024; Sargent et al. 2024) nor total (Shangguan et al. 2019) gas content of currently interacting or recently merged galaxies have found evidence for significant gas depletion.  The largely normal (or even enhanced) gas fractions in mergers apparently undermines the connection with quenching.  However, it may be that gas reservoirs remain intact, but are too turbulent to efficiently form stars (Smercina et al. 2022; Brunetti et al. 2024).  Moreover, global measurements of gas content in mergers are unable to capture the spatial diversity of the galactic response to the interaction.  High resolution maps of gas in mergers indicate a wide range of depletion times, and often a suppression of star formation (e.g. Bemis \& Wilson 2019; Tomicic et al. 2019; Thorp et al. 2022).  Ideally, we need a more direct way to assess the shutdown of star formation in mergers.

\smallskip

As tracers of recent and rapid quenching, post-starburst (PSB) galaxies offer just such an opportunity (French 2021).  PSB galaxies are characterized by a combination of spectral characteristics that simultaneously indicate recent, but not on-going, star formation (Goto 2005; Wild et al. 2007; Alatalo et al. 2016).  Although there were early indications that PSB signatures were relatively common in mergers (Liu \& Kennicutt 1995) recent observational work has robustly quantified the prevalance of recent and rapid quenching in post-mergers, finding statistical excesses by factors of 10-60 (Ellison et al. 2022; Li et al. 2023b).  Combined with the complementary observations that a high fraction of PSBs are in (post) mergers (Pawlik et al. 2016, 2018; Meusinger et al. 2017; Sazonova et al. 2021; Wilkinson et al. 2022) this indicates that mergers are able to rapidly shut down star formation (although it has been argued that \textit{most} quenching happens by other means, e.g. Bluck et al. 2016; Weigel et al. 2017).  Whilst this body of work indicates a link between interactions and quenching, it is currently unconstrained exactly \textit{when} in the merger sequence this quenching takes place.  By assessing the excess of PSB galaxies in \textsc{mummi} mergers as a function of time since coalescence, the work presented here makes the first temporal measurement of quenching in the post-merger regime.

\section{Data}\label{data_sec}

The goal of the current work is to track the timescale of rapid quenching during galaxy interactions.  In order to achieve this objective, we must assemble samples of both pre-coalescence and post-merger galaxies, as well as control samples for each.  In this section we describe the compilation of these various datasets.  Throughout this work we use redshifts taken from the Sloan Digital Sky Survey (SDSS) Data Release 7 (DR7) and stellar masses and star formation rates from the MPA-JHU catalog (Kauffmann et al. 2003; Brinchmann et al. 2004).

\subsection{The galaxy pair sample}\label{pair_sec}

The pre-coalescence phase of the merger sequence is commonly characterized by using samples of spectroscopic galaxy pairs, using the projected separation as a rough proxy for time  (e.g. Ventou et al. 2019; Pfister et al. 2020).  In the work presented here, we use the sample of wide pairs that was first compiled by Patton et al. (2016).  Briefly, the Patton et al. (2016) sample is constructed by identifying the closest companion of each galaxy in the SDSS DR7 whose stellar mass is within a factor of 10 and whose velocity separation $\Delta v < 1000$ \kms.  For the work presented here, we define a more restrictive pairs sample by selecting galaxies whose closest companions are within 100 kpc and $\Delta v < 300$ \kms\ in order to minimize contamination by projections.  Furthermore, in order to have a sample that is consistent with the post-mergers (see next sub-section) we also restrict the stellar mass range to be log ($M_{\star}/M_{\odot}) \ge 10$.  Finally, since a good quality spectral continuum will be required in order to have robust measurements of absorption line features critical for the identification of PSB features, we also require a $g$-band S/N$>$8.  There are 15,831 such galaxies that fulfill all of these criteria and that represent our pairs sample\footnote{Strictly, this is a sample of galaxies with close companions, rather than a true sample of pairs, because the closest companions identified do not always find unique pairings.  I.e. Galaxy A may have galaxy B as its closest companion, but galaxy B's closest neighbour may be galaxy C.  This explains why we have an odd number of galaxies in the `pairs' sample.  Nonetheless, we use the nomenclature `pairs' as it is widely adopted in the literature.}.

\subsection{The post-merger sample}\label{pm_sec}

In this sub-section we first review the design of the \textsc{mummi} neural network merger identifier, and then describe its application to UNIONS $r$-band imaging that yields the post-merger samples used in this work.

\subsubsection{An overview of the \textsc{mummi} merger identification pipeline}\label{mummi_sec}

\textsc{mummi} (Ferreira et al. 2024a, b) is a neural network based merger classifier that was trained on galaxies identified in the IllustrisTNG 100-1 (hereafter, TNG) simulation (Nelson et al. 2019).  The TNG training sample consists of a balanced set of mergers and non-interacting galaxies with log(M$_{\star}$/M$_{\odot}$) $>$ 10 at $z<1$.  The merger sample consists of galaxies up to 1.8 Gyr before and after coalescence, i.e. tracing both the pair and post-merger phase. In order that the neural networks are trained to optimally recognize galaxy morphologies in real data, it is important to introduce `observational realism' into the training data (e.g. Bottrell et al. 2019). The TNG mass maps are therefore convolved with a point spread function representing the median seeing in the UNIONS $r$-band imaging (described further in Section \ref{unions_sec}) and inserted into actual UNIONS frames to replicate not only sky depth and projected companions, but also imaging artifacts.   

\smallskip

\textsc{mummi} aspires to achieve three levels of classification.  First (Step 1) is a separation of isolated galaxies from pairs+post-mergers.   As mentioned in the Introduction, one of \textsc{mummi}'s novel characteristics is that it uses an ensemble of 20 networks that each gives an independent merger predition (i.e. merger or non-merger), providing the user flexibility to choose how many `votes' (out of 20) are required in order to receive a final positive merger classification.  Ferreira et al. (2024a) assess the performance of two fiducial cases: a simple majority and a unanimous case in which $>$10/20 and 20/20 networks must return a positive merger classification, respectively.  In Step 2, \textsc{mummi} separates pairs from post-mergers; for this simpler task, \textsc{mummi} uses just one pair of neural networks.

\smallskip

It is Step 3, described in Ferreira et al. (2025b), of the \textsc{mummi} pipeline that is of particular import to our study. In this final step, post-mergers found in Step 2 are classified according to their time since merger ($T_{PM}$), information that is available in the training dataset via simulation snapshot number.  Thus, the Step 3 prediction is not a continuous number in time, but rather a prediction of how many snapshots have passed since coalescence of the two sub-halos in TNG, where each snapshot is separated by 160 Myr, on average.   Put another way, Step 3 is another classification step, rather than a regression, with possible outputs between 1 and 11 snapshots since coalescence ($0 < T_{PM} < 1.76$ Gyr) in bins of width 160 Myr.

\smallskip

Furthermore, Ferreira et al. (2025b) have shown that whilst classification of snapshots soon after the merger is achieved with high performance, there is considerable confusion (uncertainty) between predictions of adjacent snapshots at intermediate and long times.  To alleviate this uncertainty Ferreira et al. (2025b) propose two additional refinements in Step 3.  First is to introduce a binning scheme in which later snapshots are combined into a single time bin.  Therefore, whilst Step 3 nominally places each galaxy into one of 11 snapshots post-merger, Ferreira et al. (2025b) propose keeping snapshot 1 as a stand-alone time bin, but combining the predictions of snapshots 2-3, 4-6 and 7-11.  These four time bins therefore represent $0<T_{PM}<0.16$, $0.16<T_{PM}<0.48$, $0.48<T_{PM}<0.96$ and $0.96<T_{PM}<1.76$ Gyr respectively.  Finally, Ferreira et al. (2025b) use the snapshot probability distributions for each galaxy to identify those with the most robust predictions, which yields an even higher quality in the time bin predictions.  

\subsubsection{Application of \textsc{mummi} to UNIONS $r$-band imaging}\label{unions_sec}

UNIONS is a collaboration of wide field imaging surveys of the northern hemisphere (Gwyn et al. in prep).  In the work presented here we make use of the UNIONS $r$-band imaging that was obtained at the 3.6-m Canada France Hawaii Telescope (CFHT).  Specifically, we make use of the fifth data release (DR5) of $r$-band imaging that covers just under 5000 $deg^2$. The seeing ranges from 0.45 to 1 arcsecond, with a median of 0.7 arcseconds. The median point-source 5$\sigma$ depth is 24.9, as measured through a 2 arcsecond aperture (Gwyn et al. in prep).

\smallskip

In order to study detailed galaxy characteristics, and how these are affected by a merger, we rely heavily on spectroscopic data to infer properties such as star formation rate, stellar mass and black hole accretion status.  We are therefore interested in the overlap region between UNIONS $r$-band and the SDSS DR7.  Ferreira et al. (2024a) identify 235,354 galaxies with $0.03 < z < 0.5$ in this overlap region and present the results of Steps 1 and 2 (merger identification and classification as pair or post-merger) of the \textsc{mummi} pipeline.  We make use of the vote outputs\footnote{Ferreira et al. (2024a) present a full catalog of galaxies and \textsc{mummi} vote outcomes that can be used to recreate the work presented here.} from Step 1 to identify post-mergers that have a majority classification, i.e $>10/20$ of the networks assign a merger probability $p(x)>0.5$. As shown in Ferreira et al. (2024a), the majority approach represents a good compromise between purity and completeness.  Since the \textsc{mummi} pipeline was trained using galaxies whose stellar mass log($M_{\star}/M_{\odot}) > 10$, we impose the same limit on the observational sample.  Our final sample, for which we have both SDSS DR7 spectroscopy with our required $g$-band S/N$>$ 8, and \textsc{mummi} merger predictions, is 134,162 galaxies of which 8128 are identified as post-mergers under the majority voting scheme.  Imposing the recommended quality threshold for time bin predictions reduces this to 5927 post-merger galaxies of which there are 559, 556, 627, 4185 galaxies in the $0 < T_{PM} <$ 0.16 Gyr, $0.16 < T_{PM} <$ 0.48 Gyr, $0.48 < T_{PM} <$ 0.96 Gyr and $0.96 < T_{PM} <$ 1.76 Gyr time bins, respectively.

\subsection{Control matching}\label{control_sec}

\begin{figure}
	\includegraphics[width=8.5cm]{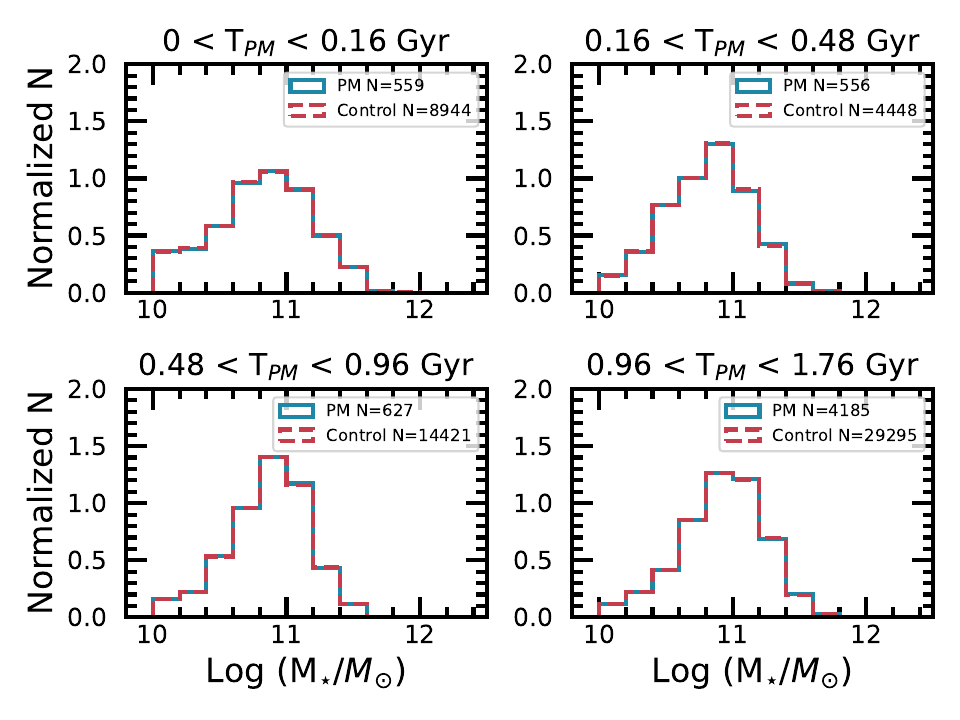}
        \caption{The stellar mass distributions of the post-merger sample in the four $T_{PM}$ time bins (blue) and matched control sample (red dashed). }
        \label{mhist}
\end{figure}

\begin{figure}
	\includegraphics[width=8.5cm]{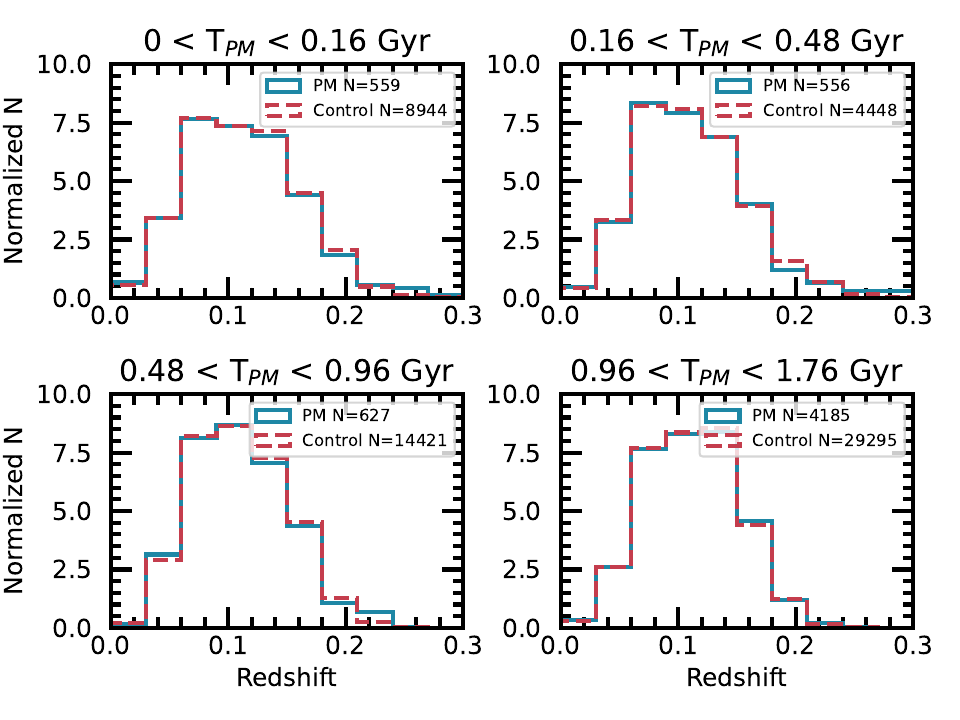}
        \caption{The redshift distributions of the post-merger sample in the four $T_{PM}$ time bins (blue) and matched control sample (red dashed). }
        \label{zhist}
\end{figure}

A control sample of non-merger galaxies is constructed for both the pairs and the post-merger samples, in order that their post-starburst fractions can be differentially assessed.  The matching procedure is run separately for the pairs and for each of the four $T_{PM}$ samples, but the same over-arching methodology is applied to all samples.

\smallskip

First, a pool of potential control galaxies is identified for both the pair and post-merger samples.  The galaxies in these control pools are essentially all galaxies that do not qualify as either pairs or post-mergers.  Specifically, for the pairs, the control pool must represent galaxies without a close companion, so we require $r_p > 100$ kpc.  Galaxies in the control pool must also have a $g$-band S/N$>$8, again to ensure that we can confidently identify PSB features.  There are 300,029 galaxies in the SDSS DR7 that fulfill this criterion and hence constitute the control pool for the pairs.

\smallskip

For the post-mergers, the control pool consists of galaxies with no more than 2/20 of the neural networks assigning a merger classification ($p>0.5$), yielding 92,415 galaxies for this control pool.  Modifications to these criteria were investigated and found not to yield significant changes to our results.

\smallskip

From these pools of possible control galaxies, we now proceed with finding a matched sample that is a close statistical match to the mergers.  For each merging galaxy in a given sample (e.g. each galaxy in the pairs sample), the control pool galaxy which is the closest simultaneous match in both log $M_{\star}$ and $z$ is identified via a \textsc{kdtree} approach and then removed from the pool. Once the best match has been found for all mergers in a given sample (e.g. the pairs), a Kolmogorov-Smirnov (KS) test is used to assess whether either the stellar mass or redshift distributions are significantly different ($p<0.99$) between the mergers and the matched controls.  If yes, the matching is halted.  If there is no statistical difference, a second round of matching is performed, taking now the second best simultaneous match in log $M_{\star}$ and $z$ and the KS test is repeated.  The matching process is repeated until the KS test fails in either the stellar mass or redshift comparison, at which point the last matching iteration is adopted.  In this way, the mergers in each sample have the same number of matched controls, but that number can be different from one sample to the next.  Also, since the matching is run separately for the pairs and the four post-merger samples, whilst a given control can not be re-used in a given sample, it can be re-used for different samples (e.g. pairs and four post-merger time bins).  The largest number of controls is achieved for the $0.48 < T_{PM} < 0.96$ Gyr bin (23) and the smallest for the $0.96 < T_{PM} <$ 1.76 Gyr bin (7).  The pairs sample achieves 8 matches per galaxy.    

\smallskip

As a demonstration of the excellent quality of the matching, we show in Figures \ref{mhist} and \ref{zhist} the normalized histogram distributions of stellar mass and redshift for the four post-merger time bins, where the post-merger distributions are shown in blue and the controls in red.  The legend reports the total number of controls identified for each sample. It can be seen that both the stellar mass and redshift distributions are indistinguishable between mergers and controls.

\subsection{Classification of post-starburst galaxies}\label{psb_sec}

The original classification of PSB galaxies used a combination of strong Balmer absorption and a lack of emission lines (e.g. Zabludoff et al. 1996; Goto 2005).  The motivation for this combination was the simultaneous presence of A and F type stars (responsible for the strong Balmer absorption) as indicators of recent star formation, but the lack of O and B stars (the very youngest populations) that generated the bulk of ionizing photons that produced HII regions.  The combination of these features led to the nomenclature of `E+A' galaxies - an elliptical-like (i.e. no emission lines) spectrum super-imposed with A star-like absorption.

\smallskip

Although the E+A selection provides a robust way to identify recently quenched galaxies, the strict cut on the presence of emission lines renders it highly incomplete.  Quenched (or quenching) galaxies may exhibit emission lines from non-stellar photoionization, such as shocks or AGN, or simply weak emission from the fading starburst.  There have therefore been several efforts to develop more inclusive selection criteria (e.g. Yan et al. 2006; Mendel et al. 2013; Yesuf et al. 2014; Alatalo et al. 2016; Meusinger et al. 2017).

\smallskip

In the work presented here we will investigate two possible PSB selection methods.  First, in order to select traditional E+A PSBs, we use the classifications in the SDSS DR7 catalog of Goto (2007) which were identified according to the following equivalent width (EW) thresholds (where positive values indicate absorption and negative values indicate emission): EW(H$\delta$)$>$ 5.0 \AA, EW([OII])$>-$2.5 \AA\ and EW(H$\alpha$)$>-$3.0 \AA.

\smallskip

Second, for a more inclusive sample of PSBs, we use the principal component analysis (PCA) of Wild et al. (2007), again applied to SDSS DR7, which selects galaxies with a relative excess of Balmer absorption given the age of their stellar population and ignores emission line strengths.  In addition to cuts in PCA space, we additionally apply mass dependent thresholds in Balmer decrement designed to remove dusty galaxies that may not be true post-starbursts.  The complete details of the PCA selection are explained in full in Wilkinson et al. (2022).

\section{Results}\label{results_sec}

Having constructed our merger and control samples and established the criteria for PSB selection, we now have all of the requisite components to investigate the connection between galaxy interactions and quenching.  Although the primary focus of this work is to investigate the frequency (and excess) of recent quenching as a function of $T_{PM}$ (Section \ref{psb_xs_sec}) we also invert the experiment in order to quantify what fraction of PSBs are mergers (Section \ref{merger_sec}).

\subsection{Quenching through the merger sequence}\label{quench_sec}

\begin{figure}
	\includegraphics[width=8.5cm]{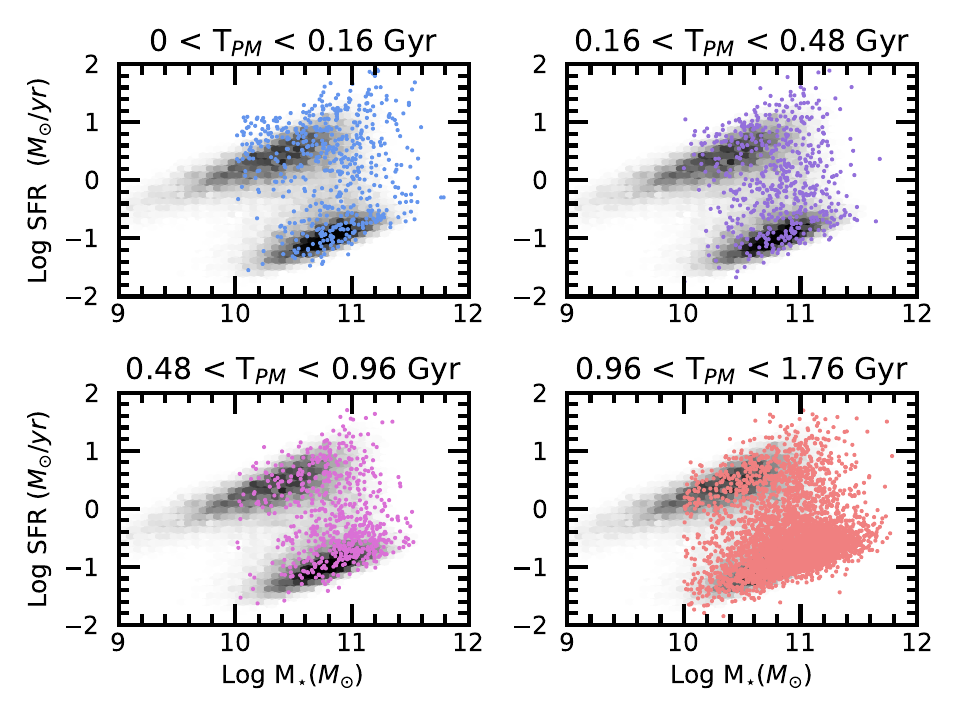}
        \caption{Star formation rate vs. stellar mass for the four $T_{PM}$ bins in our sample.  The greyscale in each panel shows the distribution of SFR and $M_{\star}$ for non-merger control pool galaxies in the UNIONS \textsc{mummi} sample.  Coloured points show the post-merger galaxies.  As time progresses since coalescence (i.e. increasing $T_{PM}$) the SFR demographics evolve from a population dominated by highly star-forming galaxies at short $T_{PM}$ towards a high fraction in the green valley at intermediate times, before quenched galaxies dominate by 1 Gyr post-merger. }
        \label{sfms_fig}
\end{figure}

We begin our investigation of quenching through the merger sequence with a top level view of SFRs in the post-merger sample.  The SFRs are taken from the MPA-JHU catalog (Brinchmann et al. 2004) wherein the H$\alpha$ emission line is used when the galaxy is star-forming and a calibration of the D4000 break otherwise (e.g. AGN dominated or quenched).  A colour-based aperture correction is applied to the SFRs derived from the fibre spectroscopy.  In Figure \ref{sfms_fig} we plot the SFR vs stellar mass of galaxies in each of the $T_{PM}$ bins.  The background greyscale in each panel shows the distribution of control pool (non-merger) galaxies for reference.  Beginning with the top left panel, whose coloured points show the SFRs for 559 galaxies with $0 < T_{PM} <$ 0.16 Gyr, we see that not only are a large fraction of post-mergers star-forming, a significant fraction are elevated above the star-forming main sequence traced out by the control pool (non-mergers), i.e. into the starburst regime.  As time progresses, we see that post-mergers appear to migrate into the intermediate SFR regime often referred to as the green valley (top right and lower left panels of Figure \ref{sfms_fig}).  In the longest time bin, $0.96 < T_{PM} < 1.76$ Gyr shown in the lower right panel of Figure \ref{sfms_fig}, the majority of post-mergers lie on the quenched sequence. 

\smallskip

Ferreira et al. (2025a) have previously shown that the SFRs of star-forming galaxies in the \textsc{mummi} identified post-mergers are elevated by a factor of 1.8 immediately following the merger, declining to `normal' values by the longest $T_{PM}$ bin.  Figure \ref{sfms_fig} complements those results by showing that the overall demographics shift from star-forming to quenched as time since coalescence increases.

\smallskip

\begin{figure}
	\includegraphics[width=8.5cm]{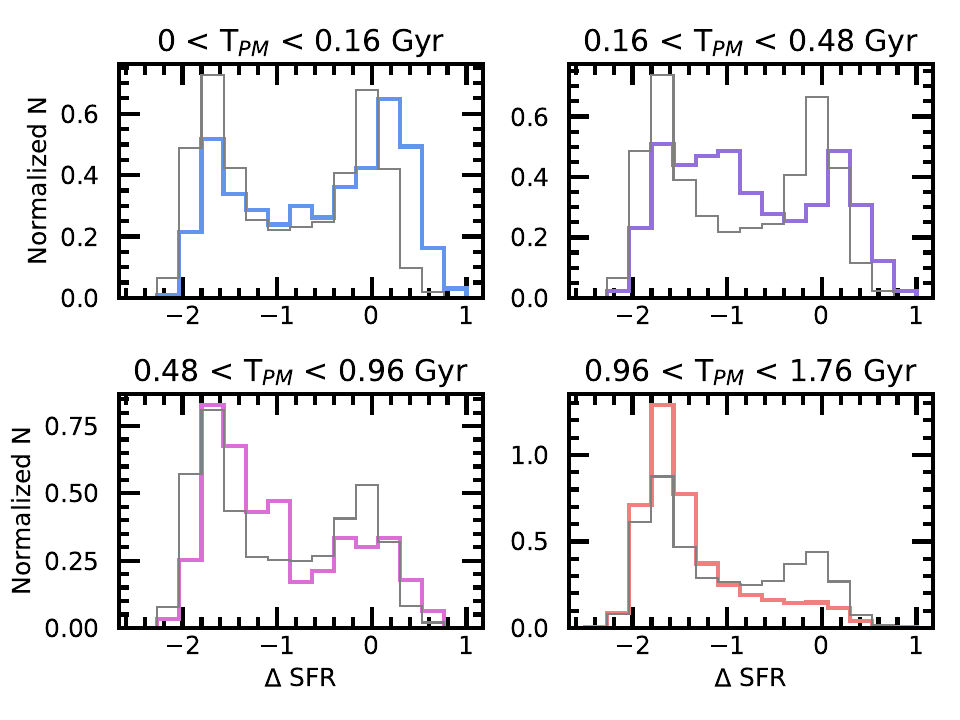}
        \caption{Distribution of $\Delta$ SFR for post-mergers in four time bins.  In each panel the grey histogram is the control sample for a given time bin and the coloured line shows the post-mergers.  At early times, post-mergers show an excess of star-forming galaxies and an elevated fraction of starbursts.  As time progresses, the post-merger becomes dominated first by green valley galaxies at intermediate times, and then by quenched galaxies by $0.98 < T_{PM} < 1.76$ Gyr. }
        \label{dsfr_fig}
\end{figure}

A more quantitative demonstration of the evolution in star-forming properties is presented in Figure \ref{dsfr_fig} where we plot the distribution of the main sequence offset, $\Delta$ SFR, of the post-merger and control galaxies in each time bin.  Following our previous work (Ellison et al. 2013; 2018b) \dsfr\ is computed for each galaxy as the difference in log SFR from a mass and redshift matched sample of star-forming galaxies (i.e. those that define the star forming main sequence).  Although the distribution of \dsfr\ is different for each of the four control samples, we see that all four grey histograms in Figure \ref{dsfr_fig} show the same general (and well-known) characteristics.  Namely, a bimodal distribution of \dsfr\ (star-forming and quenched galaxies separated by a green valley) in which the quenched population (low \dsfr) outnumbers the star-forming population.  The post-mergers deviate from the control sample in different ways as the merger progresses.  At short $T_{PM}$ (top left panel of Figure \ref{dsfr_fig}) we find that it is the star-forming population that dominates and with an elevated fraction of starburst galaxies (\dsfr\ $>$ 1). For post-mergers in the $0.16 < T_{PM} < 0.48$ time bin (top right panel in Figure \ref{dsfr_fig}) we see that the green valley ($-1.5 < \Delta$ SFR $< -0.5$) is largely filled in.  By the longest time bin (0.96 $< T_{PM} < 1.76$ Gyr, bottom right panel of Figure \ref{dsfr_fig}) we see that post-mergers are now under-abundant in star-forming galaxies, and that the quenched sequence is now highly populated.  Taken together, Figures \ref{sfms_fig} and \ref{dsfr_fig} paint a picture in which post-mergers are dominated by star-forming galaxies (with a notable fraction of starbursts) soon after coalescence, but quench their star formation over the course of some 2 Gyr.

\begin{table*}
  \begin{center}
  \caption{Statistics for PSB identification using the PCA method.  In each time bin, we tabulate the number of post-mergers, number of PSBs in the post-merger sample, the fraction of post-mergers that are PSBs and then the equivalent statistics for the control sample.  The final column shows the ratio of the fraction of PSBs in the post-mergers and controls, hence representing a PSB excess triggered by the interaction.}
\begin{tabular}{lccccccc}
\hline

$T_{PM}$ (Gyr) & $N_{PM}$ & $N_{PM,PSB}$ & $f_{PM,PSB}$ & $N_{control}$ & $N_{control,PSB}$ & $f_{control,PSB}$ & Excess \\ 

\hline                                                                                                      
0-0.16    &  559 &  65 & 0.116$\pm$0.014 &  8944 & 45 & 0.0050$\pm$0.0007 & 23.1$\pm$4.4 \\
0.16-0.48 &  556 & 103 & 0.185$\pm$0.016 &  4448 & 26 & 0.0058$\pm$0.0007 & 31.7$\pm$6.8\\
0.48-0.96 &  627 &  42 & 0.067$\pm$0.010 & 14421 & 64 & 0.0044$\pm$0.0006 & 15.1$\pm$2.9\\
0.96-1.76 & 4185 &  59 & 0.014$\pm$0.002 & 29295 &105 & 0.0036$\pm$0.0004 &  3.9$\pm$0.6\\

\hline
\end{tabular}
\label{pca_tab}
  \end{center}
\end{table*}

\begin{table*}
  \begin{center}
  \caption{As for Table \ref{pca_tab} but for PSB identification using the E+A method.  }
\begin{tabular}{lccccccc}
\hline

$T_{PM}$ (Gyr) & $N_{PM}$ & $N_{PM,PSB}$ & $f_{PM,PSB}$ & $N_{control}$ & $N_{control,PSB}$ & $f_{control,PSB}$ & Excess \\ 

\hline                                                                                                      

0-0.16    & 559   & 20 & 0.036$\pm$0.008 & 8944  & 9 & 0.0010$\pm$0.0003 & 35.6$\pm$14.2 \\
0.16-0.48 & 556   & 37 & 0.067$\pm$0.011 & 4448  & 3 & 0.0007$\pm$0.0004 & 98.7$\pm$59.1\\
0.48-0.96 & 627   &  9 & 0.014$\pm$0.005 & 14421 & 6 & 0.0004$\pm$0.0002 & 34.5$\pm$18.1\\
0.96-1.76 & 4185  &  7 & 0.002$\pm$0.001 & 29295 & 8 & 0.0003$\pm$0.0001 &  6.1$\pm$3.2\\

\hline
\end{tabular}
\label{goto_tab}
  \end{center}
\end{table*}

\subsection{What fraction of mergers are post-starbursts?}\label{psb_xs_sec}

Having established the evolution of the post-merger population from pre-dominantly star-forming towards predominantly quenched as $T_{PM}$ increases, we now turn to PSBs as specific tracers of recent quenching.  In Tables \ref{pca_tab} and \ref{goto_tab} we present the statistics of PSB frequency in the various post-merger and matched control samples, using the PCA and E+A selection techniques respectively.  Errors on all of these fractions are calculated using binomial statistics.  The control sample statistics highlight the known rarity of PSBs in `normal' galaxies - typically much less than 1 per cent of galaxies exhibit the signature of recent and rapid quenching. Indeed, when using the more restrictive E+A selection method the fraction of PSBs in the control population is $\sim$0.05 per cent.  As expected, the fraction of PCA PSBs in the controls is higher, thanks to the sensitivity of the selection method to weaker bursts and the inclusion of weak emission lines.  In contrast with the low incidence of PSBs in the control samples, Tables \ref{pca_tab} and \ref{goto_tab} show that in post-mergers the PSB fractions can reach almost 20 per cent (i.e. for PCA identification in the 0.16 $< T_{PM} < 0.48$ Gyr bin).  In the final columns of Tables \ref{pca_tab} and \ref{goto_tab} we compute the PSB excess, i.e. the fraction of mergers with PSBs divided by the fraction of control galaxies with PSBs.  These PSB excesses range from $\sim$ 4 to almost 100, depending on the PSB selection technique and the time post-merger.

\smallskip

\begin{figure}
	\includegraphics[width=8.5cm]{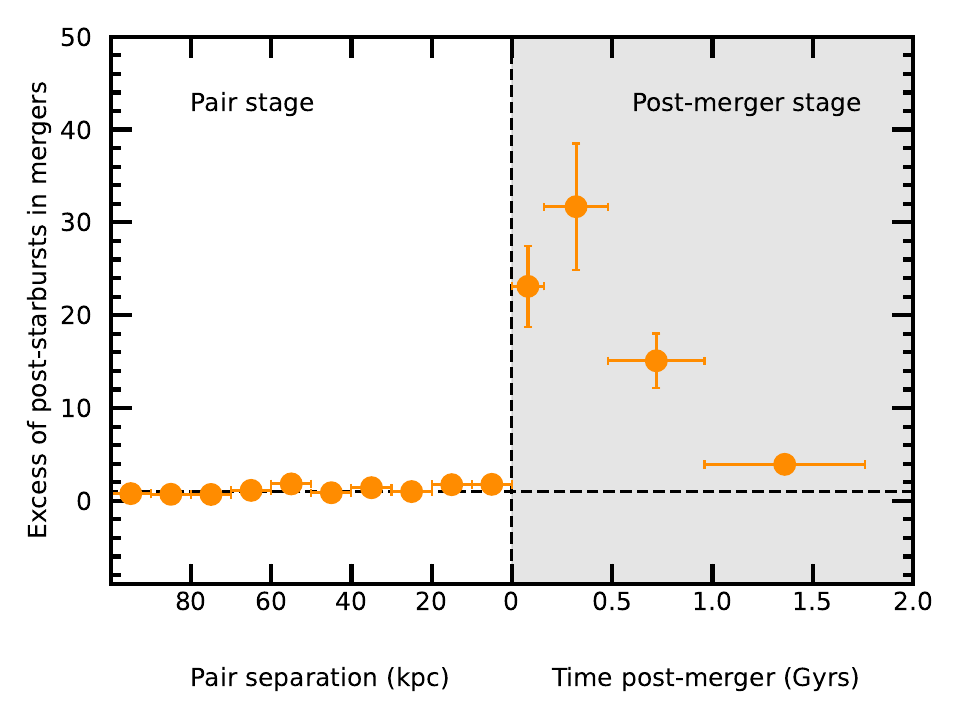}
        \caption{The excess of PCA-identified PSB galaxies in the pairs (left side of the diagram) and post-mergers (right side of the diagram) compared with the control sample. The horizontal dashed line shows an excess of one, i.e. the same PSB fraction in mergers and controls.   The pair sample is binned in units of projected separation (kpc) whereas the post-mergers are binned by the time post-merger, which in turn comes from the snapshot classification of \textsc{mummi}'s Step 3.  X-axis error bars indicate the width of the $r_p$ or $T_{PM}$ bin. }
        \label{pca_xs}
\end{figure}

\begin{figure}
	\includegraphics[width=8.5cm]{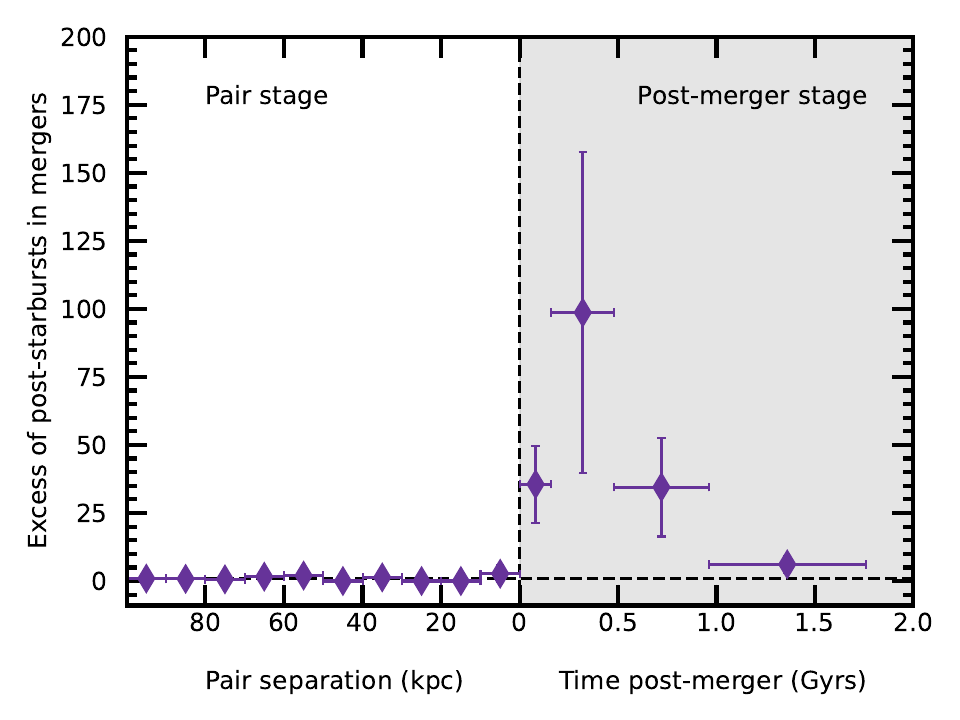}
        \caption{As for Figure \ref{pca_xs} but using the E+A method for PSB identification. }
        \label{goto_xs}
\end{figure}

Figures \ref{pca_xs} and \ref{goto_xs} present a visual summary of our results, again considering the PCA and E+A PSB selections, respectively.  The horizontal dashed line shows an excess of one, i.e. the same PSB fraction in mergers and controls.  The left side of each figure shows the PSB excess for the galaxy pairs sample, calculated in projected separation bins of 10 kpc width.  Note that pair separation is plotted decreasing from left to right, in contrast to the convention adopted in most previous works (including our own).  The rendering in Figures \ref{pca_xs} and \ref{goto_xs} is chosen so that the plot can be read from left to right as a continuous merger sequence, from the pre- to post-coalescence regimes.  On the right side of Figures \ref{pca_xs} and \ref{goto_xs} the PSB excess in the post-mergers is plotted in sequence of increasing time since merger.  We remind the reader that due to variable confusion between \textsc{mummi}'s time prediction, the $T_{PM}$ bins are not uniform, a feature shown by the error bars on the x-axis.

\smallskip

As we have previously shown in Ellison et al. (2022), Figures \ref{pca_xs} and \ref{goto_xs} demonstrate that, although some PSBs are found in pre-coalescence pairs (e.g. Baron et al. 2024) there is minimal statistical enhancement compared with the control sample in the pair regime.  The agreement in the pair regime with our previous work is unsurprising since we are using the same sample as previously published in Ellison et al. (2022), the only difference being a slightly different matching scheme (mass and redshift, as opposed to only mass in Ellison et al. 2022).  In all $r_p$ bins the uncertainty (computed by propagating the binomial errors on each fractional uncertainty) on PSB excess in pairs is consistent with a value of one (i.e. no excess). There is a very tentative hint of a PSB excess in the smallest $r_p$ bins (again, also seen in Ellison et al. 2022); a factor of 1.75$\pm$0.98 for the PCA selection and 2.67$\pm$3.08 for the E+A selection, but these are not statistically significant given the error bars. If we increase the $r_p$ bin width to 20 kpc, the E+A excess drops, brought down by a lower PSB fraction in the 10-20 kpc bin, to 0.80$\pm$0.84.  However, the PCA excess in the smallest $r_p$ bin becomes slightly more robust with an excess of 1.75$\pm$0.48, the first (albeit marginally) statistically significant detection of an increase in PSBs in the close pair ($<20$ kpc) regime.

\smallskip

In contrast to the pair regime, Figures \ref{pca_xs} and \ref{goto_xs} show a dramatic enhancement in the PSB fraction in the post-merger regime. Qualitatively the picture is the same for both the PCA (Figure \ref{pca_xs}) and E+A  (Figure \ref{goto_xs}) selection methods - an increase in the PSB fraction begins soon after coalescence (our first time bin captures $0< T_{PM} < 0.16$ Gyr), continuing to rise (and peaking) over the course of the next $\sim$ 300 Myr.  In the longest time bin ($0.96 < T_{PM}$ < 1.76 Gyr) the excess is much decreased, but still measured at a statistically significant level.

\smallskip

Quantitatively, the PCA and E+A methods give quite different results, with PSB excesses that peak at $\sim \times 30$ for the former and $\sim \times 100$ for the latter.  Changing the threshold number of votes required in \textsc{mummi}'s Step 1 does not significantly alter these numbers.  Even with a unanimous approach (all 20 of the networks required to identify a given galaxy as a merger in Step 1) the excesses are in good agreement with those shown in Figures \ref{pca_xs} and \ref{goto_xs} (albeit with larger error bars because of the smaller samples).

\smallskip

For comparison with the results presented in Figures \ref{pca_xs} and \ref{goto_xs}, in Ellison et al. (2022), where we considered a smaller sample (699 post-mergers taken from the UNIONS DR2) with no $T_{PM}$ information we found an excess of $\sim$ $\times30$ and $\times60$ for the PCA and E+A methods respectively.  Since the post-merger sample presented in Ellison et al. (2022) was selected using a CNN trained on the first TNG simulation snapshot after coalescence (Bickley et al. 2021) it is perhaps unsurprising that the PSB excesses measured therein are an average of the first few time bins shown in Figures \ref{pca_xs} and \ref{goto_xs}.  However, the ability to split the mergers into 160 Myr time bins has revealed a more detailed sequential picture of rapid quenching in the post-merger regime.  

\subsection{What fraction of post-starbursts are mergers?}\label{merger_sec}

In the previous section we quantified what fraction of mergers are PSBs; our results showed that there is a factor 30-100 more PSBs in recent ($0.16 < T_{PM}<$ 0.48 Gyr) post-mergers compared with expectations from a control sample.   In this section we now turn to the inverse question: what fraction of PSBs are in mergers?  Whereas in the last section we demonstrated that mergers can lead to rapid quenching, the question posed in this section assesses whether mergers are the dominant source of rapid quenching.  It is necessary to ask both of these complementary questions in order to fully understand the relation between galaxy interactions and rapid quenching.

\smallskip

We begin our measurement of the PSB merger fraction by defining the PSB sample itself.  As we have done in the previous analysis presented thus far, we use two complementary selection techniques, the traditional E+A selection as well as the more inclusive PCA technique (Section \ref{psb_sec}).  We emphasize again that in order to be selected by the E+A technique a significant starburst, followed by complete quenching, must have occurred.  On the other hand, the PCA selection is sensitive to more subtle enhancements/reductions in star formation (e.g. Pawlik et al. 2018; 2019) and incomplete quenching.  These two selection techniques are therefore sensitive to different star formation histories (and potential quenching mechanisms) that make a comparative study informative.

\smallskip

Our starting point for defining the PSB samples is again the overlap between the \textsc{mummi} UNIONS DR5 catalog and SDSS DR7, including only galaxies with a $g$-band S/N$>$8 to ensure sufficient quality for spectral measurements.  Since the \textsc{mummi} networks were trained on galaxies with log ($M_{\star}/M_{\odot}) > 10$, and this is the mass regime where we expect optimal merger identification, we also impose this mass limit on the parent sample.  There are 208 and 1046 PSBs thus identified using the E+A and PCA methods, respectively.

\smallskip

\begin{figure}
	\includegraphics[width=8.5cm]{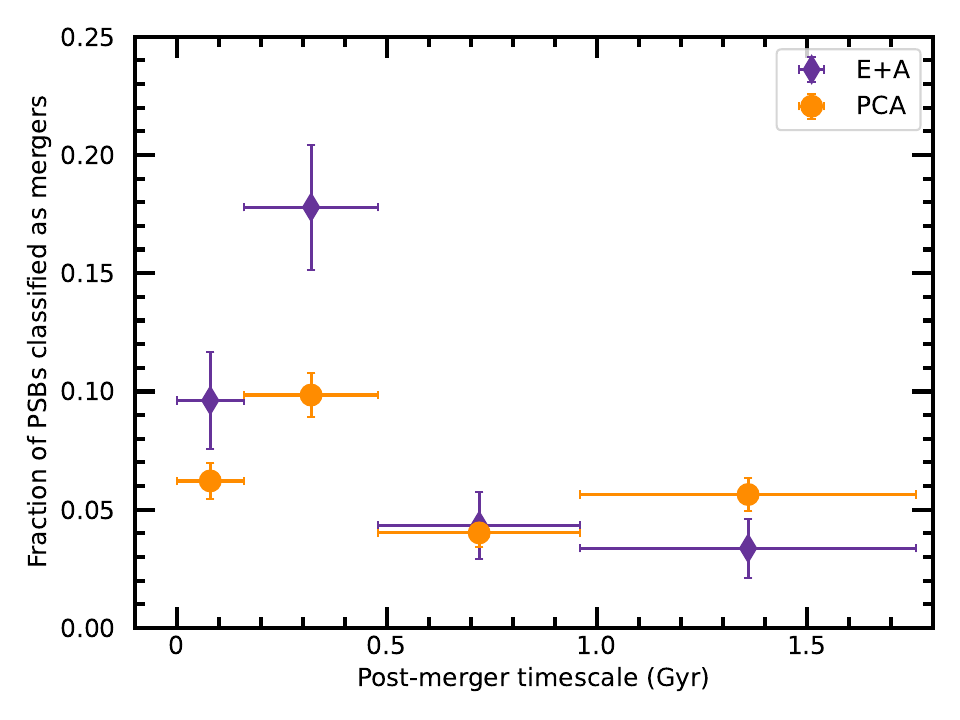}
        \caption{The fraction of PSBs classified as mergers in different time bins since the merger. Results are show for PSBs classified either via the PCA (orange) or E+A (purple) techniques.  The majority of the PSBs classified as post-mergers are in the $0.16 < T_{PM} < 0.48$ Gyr time bin.  However, these fractions are an incomplete accounting of the mergers in PSBs because only 73 per cent of the post-merger sample can be accurately placed into a time bin.  The fractions also do not account for the pre-coalescence (pair) phase.  }
        \label{pm_frac_bins}
\end{figure}

In Figure \ref{pm_frac_bins} we plot the fraction of PSBs that are classified as a post-merger as a function of $T_{PM}$.  The total fraction of PSBs found to be in post-mergers of all time bins is 26 and 35 per cent for the PCA and E+A categories, respectively.  Of the PSBs found to be post-mergers, the majority are found to be in the $0.16 < T_{PM} < 0.48$ Gyr time bin.  Figure \ref{pm_frac_bins} therefore strengthens the connection between rapid quenching and recent coalescence found in the previous section.  Our results also agree with the combined findings of Ferreira et al. (2025a) and Pawlik et al. (2016) who found, respectively, that the star formation rate in mergers peaks around coalescence and the fraction of PSBs with merger features declines with the age of the burst.

\smallskip

However, Figure \ref{pm_frac_bins} represents an incomplete accounting of the total fraction of mergers in PSBs for two reasons.  First, Figure \ref{pm_frac_bins} does not account for PSBs found to be in the pre-coalescence (pairs) phase.  Second, as described in Section \ref{unions_sec}, only 5927/8128 (73 per cent) of the post-mergers identified in Step 2 of \textsc{mummi} can be confidently placed into a time bin.  A full accounting of the merger fraction in PSBs therefore requires us to make two adjustments to our experiment.  First, we must account for the fraction of PSBs in the pre-coalescence phase.  Although in Section \ref{pair_sec} we have already defined a galaxy pair sample, this is not ideal for the measurement of the pre-merger fraction of PSBs due to the high spectroscopic incompleteness of the SDSS at small separations (Patton \& Atfield 2008).  We therefore use \textsc{mummi} for the identification of both the pre- and post-merger regimes, an assignment done in Step 2 of the pipeline (Ferreira et al. 2024a).  Second, we must relax the requirement of a high confidence in the $T_{PM}$ assignment for post-mergers used in the previous section, because we aim to calculate the fraction of PSBs in \textit{all} post-mergers, not just the ones for which we judge we can accurately predict time since coalescence.  The post-merger sample therefore increases from 5927 to 8128 galaxies.

\smallskip

\begin{figure}
	\includegraphics[width=8.5cm]{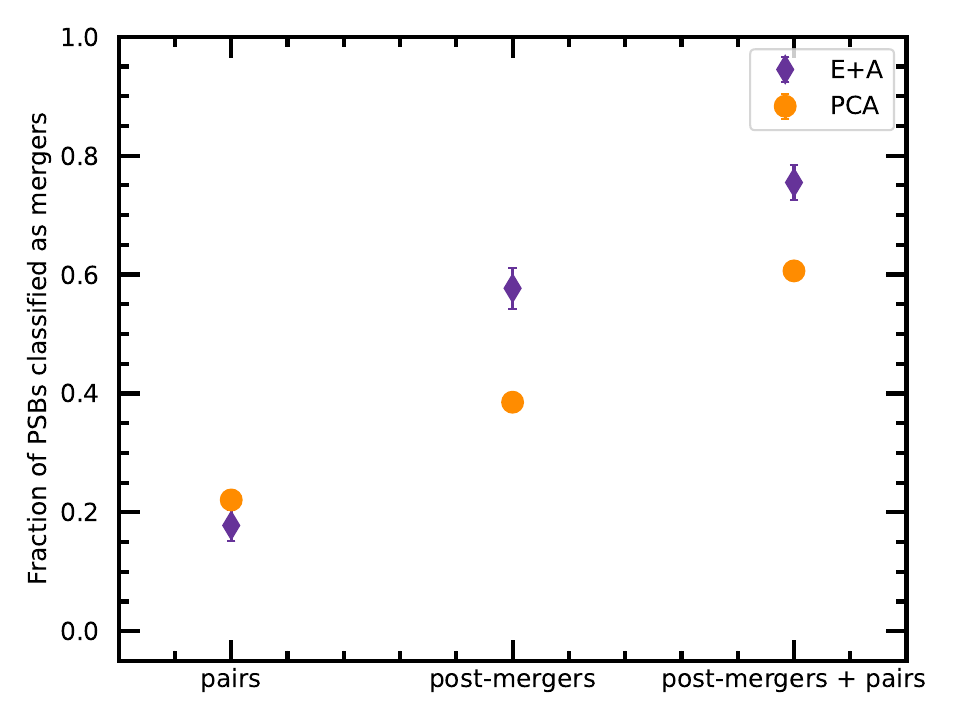}
        \caption{The fraction of PSBs classified as mergers, in both the pre-coalescence (pair) and post-merger regimes.  The sum of these fractions yields the total fraction of PSBs associated with the interaction sequence. Results are show for PSBs classified either via the PCA (orange) or E+A (purple) techniques.   }
        \label{pm_frac}
\end{figure}

In Figure \ref{pm_frac} we present the fraction of PSBs identified as pairs, post-mergers and the sum of both phases.  Following the convention of previous figures, results for PCA selected PSBs are shown in orange and E+A PSBs in purple.  Compared with Figure \ref{pm_frac_bins}, Figure \ref{pm_frac} shows that the fraction of PSBs in post-mergers has increased from 26 to 39 per cent (PCA selection) and from 35 to 58 per cent (E+A selection) once the full post-merger sample is considered.  Figure \ref{pm_frac} also shows that PSBs are more common in post-mergers than pairs, regardless of which method is used to define the PSB sample.  Specifically, 18 and 58 per cent of E+A PSBs are in pairs and post-mergers, respectively.  For PSBs selected using the PCA method, the pair and post-merger fractions are 22 and 39 per cent.  The qualitative predominance of post-mergers over pairs is consistent with results in the previous section, in which we found that galaxies in pairs have barely elevated PSB fractions, whereas post-mergers have one to two orders of magnitude more PSBs than expected from the control samples\footnote{The actual number of pairs, however, changes because in the previous section we considered only spectroscopic pairs (highly incomplete due to fibre collisions), compared with image based classifications in this section.}  Nonetheless, a significant fraction of PSBs do occur in the pre-coalescence regime of the interaction and by summing the pair and post-merger fractions we find that 75 per cent of E+A PSBs and 61 per cent of PCA PSBs are found in one of these two phases.  I.e., \textit{the majority of PSBs in our sample are linked to galaxy-galaxy interactions.}  

\smallskip

\begin{figure}
	\includegraphics[width=8.5cm]{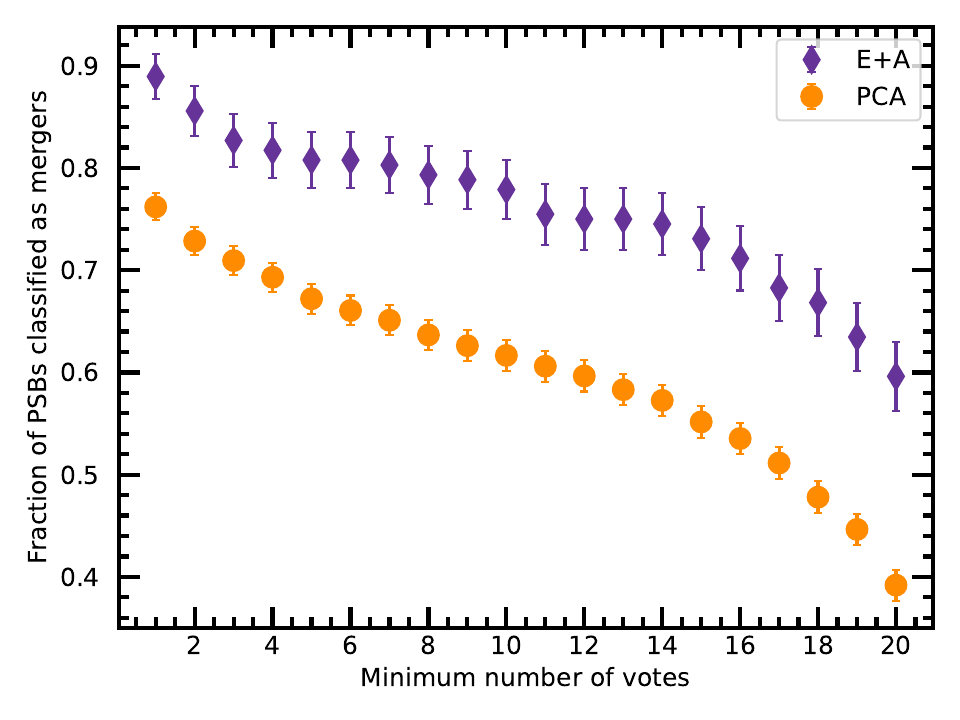}
        \caption{The fraction of PSBs classified as mergers (including both pair and post-merger classes) as a function of the minimum number of votes in \textsc{mummi}'s Step 1.  The fiducial classifications used in the work presented here require a majority vote of $>$10 votes.  Results are show for PSBs classified either via the PCA (orange) or E+A (purple) techniques.  Even with the strictest requirement of a unanimous classification (all 20 independent networks classify the galaxy as either a pair or post-merger) $>$ 60 per cent of E+A PSBs are in a merger, indicating that these line-free, rapidly quenched galaxies are dominated by mergers.  }
        \label{pm_nvotes}
\end{figure}

In the work presented so far, we have required that $>$10 out of the 20 networks in the \textsc{mummi} ensemble label a given galaxy as a merger (pair or post-merger) in Step 1.  Ferreira et al. (2024a) show that this majority voting threshold offers a good trade-off between completeness and purity and, as discussed in the previous section, the excess of PSBs in post-mergers (Figures \ref{pca_xs} and \ref{goto_xs}) are not sensitive to increasing this confidence threshold.  However, in the assessment of the fraction of mergers amongst PSBs, the choice of vote threshold is expected to have a significant effect -- we expect higher merger fractions as the vote threshold is dropped and the sample becomes more inclusive.  Likewise, as we increase the vote threshold, the merger fraction is expected to drop as we require the \textsc{mummi} ensemble to be more confident, but at the expense of completeness.  In Figure \ref{pm_nvotes} we assess the impact of the choice of vote threshold on our results by plotting the merger fraction (pairs plus post-mergers) in the UNIONS sample as a function of the minimum number of votes required for a positive classification.  The experiment is conducted for both PSBs classified through the PCA (orange data points) and E+A (purple data points) methods.

\smallskip

Figure \ref{pm_nvotes} shows that the fraction of PSBs found in mergers is very sensitive to how the PSB sample is selected.  The fraction of E+A PSBs that are identifed as mergers is uniformly higher than the fraction of mergers in the PCA PSBs, typically by about 50 per cent.  This was already seen for the majority merger selection described above (Figure \ref{pm_frac}), in which $\sim$ 60 per cent of PCA-selected PSBs with at least 10/20 merger votes, compared with 75 per cent mergers in E+A-selected PSBs.  Figure \ref{pm_nvotes} shows that this trend is stable across all vote thresholds; while the majority of E+A PSBs are mergers across our full range of voting thresholds, PCA-selected PSBs may have up to 60 per cent contribution from non-mergers for the strictest criteria.  

\smallskip

At this point, it is sage to pause to consider how complete our merger identifications are. For example, if \textsc{mummi} misses a significant fraction of mergers, perhaps the merger fraction (post-mergers + pairs) that we have found to be 75 per cent for E+A selected PSBs, could approach 100 per cent.  Such a correction could lead us to conclude that \textit{all} PSBs are in mergers, as opposed to our current conclusion that simply the \textit{majority} of PSBs are in mergers.   We assess the robustness of our computed merger fractions in two ways. First, we compute a theoretical correction factor that accounts for \textsc{mummi}'s purity ($P$) and completeness ($C$).  The true merger fraction ($f_{true}$) is calculated\footnote{A derivation of equation \ref{eqn_corr} is given in the Appendix.} from the measured merger fraction ($f_{meas}$) as

\begin{equation}\label{eqn_corr}
  f_{true} = f_{meas} \times \frac{P}{C}.
  \end{equation}

Since \textsc{mummi}'s performance (i.e. $P$ and $C$) depends on $T_{PM}$ (Ferreira et al. 2024a) $f_{true}$ must be computed for each time bin and then summed.  Because $P>C$ in some time bins and $P<C$ in other time bins (see left panel of Figure 6 in Ferreira et al. 2024a), the final value of $f_{true}$ computed turns out to be indistinguishable from the measured fraction.  For example, above we showed that 58 per cent of E+A PSBs are in post-mergers (Figure \ref{pm_frac}); the exact same fraction is recovered after accounting for purity and completeness of the \textsc{mummi} model.

As a second assessment of a potential under-estimate of the PSB merger fraction, we visually inspect the images of PSB galaxies and compare our assessment with that of \textsc{mummi}.  Since the sample of E+A PSBs is small (208 galaxies) this is an emminently manageable task.  We find that we almost always agree with the \textsc{mummi} classification of post-mergers.  Importantly, if \textsc{mummi} has classified a given PSB as a non-merger, our visual assessment agrees.  I.e. a visual assessment does not reveal `missed mergers' that would significantly increase the PSB merger fractions computed thus far.

\smallskip

\begin{figure}
	\includegraphics[width=8.5cm]{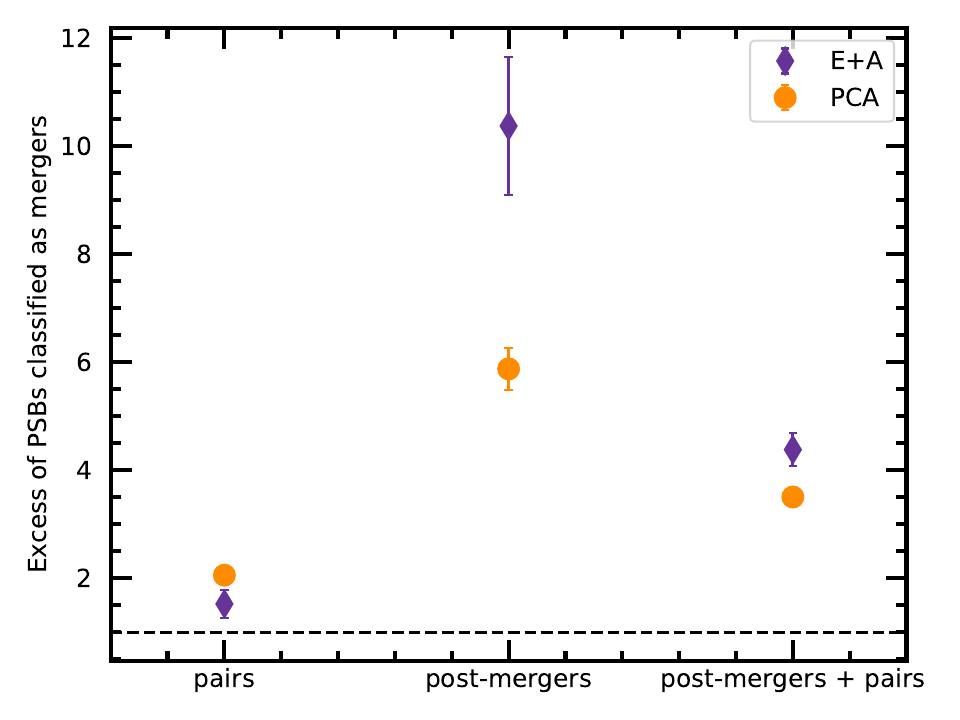}
        \caption{The excess of PSBs classified as mergers compared with a control sample.  Results are shown for PSBs classified either via the PCA (orange) or E+A (purple) techniques.   }
        \label{pm_xs}
\end{figure}

Having shown that, with some caveats on how they are selected and how mergers are identified, the majority of PSBs are linked to galaxy-galaxy interactions (e.g. 75 per cent for majority-vote selected E+A PSBs), we now quantify how much more frequently they occur compared with a control sample.  Our control matching methodology is analogous to that described in Section \ref{control_sec}.  From a control sample of non-PSB galaxies we identify the closest match to each PSB in stellar mass and redshift, repeating the process until a KS test fails.  For the PCA and E+A PSBs we find 5 and 7 non-PSB controls per galaxy, leading to control samples of 5230 and 1456 non-PSBs, respectively.  The merger fractions in these non-PSB control samples are assessed identically to the PSB samples (Figure \ref{pm_frac}), and the ratio of the fractions is shown in Figure \ref{pm_xs}.  Galaxy pairs are found to be $\sim$ 1.5 to 2 times more common in PSBs than expected from the control sample.  Therefore, whilst not dominating the PSB population (Figure \ref{pm_frac}) some rapid quenching is clearly initiated already in the pre-merger regime.   Figure \ref{pm_xs} also highlights how the relative excess of mergers in the PSB population depends not only on how the PSB sample is defined (e.g. Figure \ref{pm_nvotes}) but also how mergers are defined (i.e. interacting pair or post-merger), underlining the complexity of comparing results from different surveys (e.g. Table 1 of Wilkinson et al. 2022).

\section{Discussion}\label{discuss_sec}

\subsection{Fast quenching in mergers}

Understanding how galaxies quench has become a major theme in extra-galactic astronomy in the last decade.  Early observational work suggested that star formation shuts down via both fast and slow pathways (Schawinski et al. 2014), an idea that is supported in modern cosmological simulations (Rodriguez-Montero et al. 2019; Wright et al. 2019; Walters, Woo \& Ellison 2022).  The cohesive picture now emerging from this joint offensive of both observations (Bluck et al. 2016, 2020; Teimoorinia, Bluck \& Ellison 2016; Terrazas et al. 2016) and simulations (Piotrowska et al. 2022; Ward et al. 2022; Bluck, Piotrowska \& Maiolino 2023) is that slow quenching at low redshift is due to `preventative mode' AGN feedback, in which the accumulated energy in the galactic halo eventually starves the ISM of new fuel.  However, it is the fast quenching route on which we focus in this paper, since PSB selection identifies specific spectroscopic features sensitive to star formation that has ceased in the last 0.5 -- 1 Gyr.  Although our study focuses on the local universe, the existence of quenched galaxies at $z\sim5$ (Carnall et al. 2023) indicates that fast quenching is highly relevant at high redshifts.

\smallskip

Previous works have identified the prevalence of PSBs in mergers (Liu \& Kennicutt 1995; Ellison et al. 2022; Li et al. 2023b), demonstrating that rapid quenching can be triggered in galaxy-galaxy interactions.  The primary new insight from the work presented here is that the occurence of PSBs peaks at $T_{PM} \sim$ 500 Myr.  This is consistent with the results presented in Ferreira et al. (2025a), in which the SFR enhancement is found to peak around the time of coalescence ($T_{PM} \sim 0$).  Given that both the E+A and PCA methods select galaxies that have had starbursts $\sim$ 0.3 -- 0.8 Gyr ago (confirmed by fits of star formation histories in our targets, Wild et al. in prep) the coalescence era star formation see by Ferreira et al. (2025a) manifests as a peak in PSBs in our second time bin.  Ferreira et al. (2025a) also find that SFR enhancements persist until $\sim$ 1 Gyr post-merger, indicating that the triggering of some starbursts is delayed (or lasts) until well into the post-coalescence regime.  Again, this is consistent with our finding of a long-lived PSB excess, since galaxies that quench following a starburst at $0.5< T_{PM} < 1.0 $ Gyr will be seen as PSBs in our two longer time scale bins. Within our current framework we are not able to identify mergers with $T_{PM} > 1.76$ Gyr.  However, based on the returning to normal SFRs by $T_{PM} \sim$ 1 Gyr seen by Ferreira et al. (2025a), we would expect that the PSB fraction will also return to normal by $T_{PM} \sim$ 2 Gyr. 

\smallskip

In theory, timing the onset of quenching can help us identify the physical mechanism that causes it.  Coalescence is the regime at which both star formation (Ellison et al. 2013; Bickley et al. 2023; Ferreira et al. 2025a) and AGN activity (Bickley et al. 2024a; Ellison et al. 2025) are peaking. It might therefore be tempting to invoke feedback-driven galactic scale winds that lead to gas blowout as a potential quenching mechanism (Hopkins et al. 2006).  However, although PSBs have been shown to have outflows (Tremonti et al. 2007; Maltby et al. 2019; Baron et al. 2022) Sun et al. (2024) have shown that these are rarely able to escape the galactic ISM.   Moreover, outflows in mergers do not seem to be notably enhanced compared with normal star-forming, or AGN-hosting, galaxies (Matzko et al. 2022).   Perhaps most relevantly, post-merger galaxies seem to still be in possession of adundant gas reservoirs (Ellison et al. 2015, 2018a; Sargent et al. 2024), and the same is true for some PSBs at both low (French et al. 2015; Rowlands et al. 2015; Yesuf \& Ho 2020) and high (Suess et al. 2017; Belli et al. 2021; Bezanson et al. 2022; Wu et al. 2023) redshift.  Even though the gas fraction of PSBs does appear to decrease with time since burst (French et al. 2018; Li et al. 2019), this impact on the galactic ISM is happening after quenching has begun, indicating that any drop in gas fraction is happening as a result of merger/PSB related processes rather than the quenching being caused by the lack of gas.  We conclude that removal of gas is unlikely to be the cause of quenching seen in our post-merger sample.  Instead, we favour a scenario in which turbulence inhibits star formation.  By identifying PSB analogs in the EAGLE simulation Davis et al. (2019) found that the vast majority of recently quenched galaxies have disturbed kinematics, with observations confirming turbulent interstellar media and gas motions that are misaligned from stars (Chen et al. 2019; Otter et al. 2022; Smercina et al. 2022; French et al. 2023).


\subsection{What fraction of PSBs are mergers?  A poorly defined question.}

In Section \ref{merger_sec} we tackled the complementary (and oft-posed) question of what fraction of PSBs are found in mergers.  Statistics of the PSB merger fraction in the literature vary considerably ranging from as little as 6 per cent up to 60 per cent (e.g. Zabludoff et al. 1996; Blake et al. 2004; Goto 2005; Pracy et al. 2009).  However, since these early works were typically based on small samples, and given the intrinsic rarity of PSBs, the merger fractions were quite uncertain, preventing a meaningful dissection of the cause(s) behind the diverse results.  In the most comprehensive work to date, Wilkinson et al. (2022) addressed both the issue of sample size, and the need for a control sample finding (via visual inspection) 42 and 28 per cent of E+A and PCA selected PSBs demonstrated merger features, indicating that the merger fraction may depend on how the PSBs themselves are selected.

\smallskip

Our work, which finds elevated merger fractions in PSB galaxies, supports the finding of Wilkinson et al. (2022), but with a factor of two larger sample.  Like them, we have used two complimentary selection techniques throughout the work presented here, in order to assess just how our results depend on sample selection.  On the one hand, the E+A selection is strict, but incomplete.  On the other hand, the PCA method is more comprehensive, but includes potentially weak bursts and incomplete quenching.   We have found that, in all of our tests, the merger fraction is higher in E+A selected PSBs than in PCA selected PSBs (e.g. Figure \ref{pm_nvotes}).  This implies that the broader characteristics of PCA PSBs (i.e. permitting emission lines) are more frequently present in secular quenching.  In turn, our results suggest that merger-induced quenching may lead to ISM properties (bigger bursts, complete quenching) that differ from quenching through non-merging channels.  Whilst the existence of multiple pathways to quenching (e.g. merging vs. secular) have long been acknowledged (Schawinski et al. 2014; Davis et al. 2019; Pawlik et al. 2019), our results further suggest that the actual mechanisms and physics associated with these pathways may also differ.

\smallskip

Beyond the PSB selection itself, we have found that other issues also impact the quantification of PSB merger fraction.  For example, we have shown that the fraction of mergers in PSBs depends on the interaction stage.  Post-mergers are more frequent than pairs amongst PSBs (Figure \ref{pm_frac}), and within the post-mergers, there is a propensity towards timescales of $0.16 < T_{PM} < 0.48$ Gyr (Figure \ref{pm_frac_bins}).  It is therefore to be expected that studies that assess the merger fraction of PSBs will get very different answers if they are searching for close companions, vs. single disturbed galaxies.  Merger fractions will also depend on the depth of imaging (Wilkinson et al. 2024).  On the one hand, the merger fraction of PSBs will certainly increase as improved image quality is used (Sazonova et al. 2021), although Wilkinson et al. (2024) have shown that even with `ideal' quality imaging the merger recovery rate remains incomplete.  However, the fact that most PSBs are in recent post-mergers where tidal features tend to still be bright and distinct means that the majority of PSB mergers can be found in moderate depth imaging (Bickley et al., 2024b).

\begin{figure}
	\includegraphics[width=8.5cm]{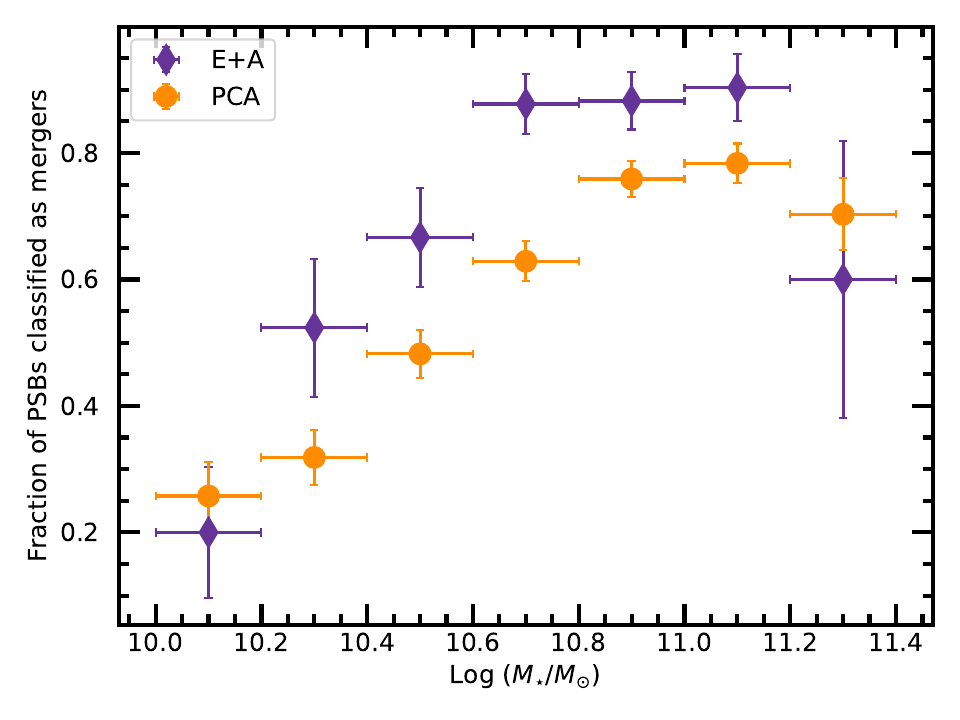}
        \caption{The fraction of PSBs classified as mergers compared with a control sample as a function of stellar mass.  Results are show for PSBs classified either via the PCA (orange) or E+A (purple) techniques.   }
        \label{pm_xs_mass}
\end{figure}

\smallskip

Finally, in Figure \ref{pm_xs_mass}, we show that the fraction of PSBs identified as mergers (pairs or post-mergers) is very sensitive to the stellar mass regime of the sample.  At low stellar masses (log $(M_{\star} / M_{\odot}) \sim 10$) the fraction of PSBs located within a merging galaxy is small, only around 20 per cent.  The merger fraction increases with increasing stellar mass, saturating at $\sim$ 90 per cent for log $(M_{\star} / M_{\odot}) > 10.7$.  A very similar result was presented by Pawlik et al. (2018), although they only considered two mass bins whereas our larger sample permits a more detailed assessment.   The mass dependence shown in Figure \ref{pm_xs_mass} is not expected to be linked to \textsc{mummi}'s capacity to find mergers at low masses; Ferreira et al (2024a) have shown that performance is stable with both mass and mass ratio.

\smallskip

Assuming that there is indeed no bias in our ability to detect mergers as a function of mass, there are two possible interpretations of the results in Figure \ref{pm_xs_mass}.  First, that mergers contribute to rapid\footnote{There may be additional slow quenching mechanisms, as described above, but since PSBs select rapid and recently quenched galaxies it is the fast mode that is relevant to our discussion here.} quenching across the full mass range, but that at low masses an alternative mechanism is significantly more effective.  Alternatively, the relative paucity of interaction features amongst PSBs at low masses may be due to mergers being ineffective at rapid quenching below log $(M_{\star} / M_{\odot}) \sim 10.4$.  These two scenarios are subtly different; the first scenario speaks to a non-merger mechanism \textit{gaining} importance at low masses, whereas the alternative is the \textit{inefficiency} of quenching in mergers in the low mass regime.  The former of these two seems more plausible and is consistent with the work presented by Pawlik et al (2018) who suggest that whilst low mass PSBs can be triggered by non-mergers, the vast majority of high mass systems have their quenching triggered by a merger (see also Rowlands et al. 2018).  Although the work presented here is unable to inform us further about the non-merger mechanism(s) that might supplement interaction driven quenching at low masses, AGN may be a viable culprit.  For example, it has been shown that while AGN have little effect on the galactic gas contents in massive galaxies (e.g. Fabello et al. 2011; Koss et al. 2021), in the low mass regime there is evidence for gas depletion (Bradford et al. 2018; Ellison et al. 2019b).  However, outflows tend to be rarer at low mass (Roberts-Borsani \& Saintonge 2019; Sun et al. 2024), so the process responsible for depleting gas fractions maybe be one of heating/ionization/molecular dissociation rather than removal (e.g. Izumi et al. 2020; Ellison et al. 2021).

\smallskip

So, what fraction of PSBs are in mergers?  The answer: it depends on many things, including the depth of imaging, definition of `PSB', definition of `merger' and the mass regime of the sample.  The discussion above should therefore act as a cautionary tale for attempts to compare different works in the literature.  Nonetheless, we have demonstrated that under certain conditions, e.g. selecting PSBs with the E+A method, and restricting to masses above log($M_{\star}/M_{\odot}) > 10.6$, and counting mergers across a wide range of pre- and post-coalescence times, the vast majority of these recently and rapidly quenched galaxies did so during an interaction.

\section{Summary}\label{summary_sec}

Using our new multi-model merger identifier (\textsc{mummi}, Ferreira et al. 2024a,b) we are able to not only identify a large and pure sample of post-mergers from UNIONS, but additionally label each galaxy with a predicted time-since-merger.  In a series of papers, we therefore present the first-ever observational studies of the the evolution of galaxy characteristics (such as star formation rate and nuclear accretion) as a function of $T_{PM}$.  In Ferreira et al. (2025a) we showed that the peak of merger-induced star formation occurs around the time of coalescence, return to `normal' values by $\sim$ 1 Gyr post-merger.  In the work presented here, we use the statistics of post-starburst galaxies in a sample of $\sim$ 16,000 spectroscopic galaxy pairs and $\sim$ 6000 post-mergers identified in UNIONS $r$-band imaging to investigate quenching along the merger sequence.  Our main conclusions are as follows.

\begin{itemize}

\item Immediately after coalescence, the post-merger population is dominated by star-forming galaxies, many of which are located above the star forming main sequence.  After a few hundred Myr post-merger, galaxies begin to migrate through the green valley.  By $T_{PM} \sim$ 1-1.5 Gyr, the majority of post-mergers are quenched, see Figures \ref{sfms_fig} and \ref{dsfr_fig}.

\item  Compared with a stellar mass and redshift matched control sample, the fraction of PSBs in post-mergers is statistically enhanced until at least $\sim$ 1.5 Gyr (the longest timescale probed in our study).  This is true for both PCA-selected PSBs (Figure \ref{pca_xs}) as well as the more traditionally selected E+A galaxies (Figure \ref{goto_xs}).  There is a clear peak in the excess of PSBs at $0.16< T_{PM} < 0.48$ Gyr, with a magnitude of $\times 30$ and  $\times 100$ for PCA and E+A selected PSBs respectively.  Since PSB selection methods identify galaxies that have rapidly quenched within the last 0.5 - 1 Gyr, our results indicate that quenching was likely initiated close to the time of coalescence.

\item  We also quantify the fraction of PSBs that are mergers and find that the majority (75 per cent) of classically selected E+A are identified as either pairs or post-mergers, with a lower merger fraction (61 per cent) amongst PCA selected PSBs.  However, the fraction of PSBs identified as mergers depends sensitively on whether we consider the pre- or post-merger phase, whether PSBs are selected using the PCA or E+A method and the mass range of the sample (Figures \ref{pm_frac} and \ref{pm_xs_mass}).  The sensitivity to these various choices may help to reconcile apparent inconsistencies reported in the previous literature.

\end{itemize}

Combined with the results of Ferreira et al. (2025a), our results paint a picture of interaction-triggered star formation that peaks around the time of coalescence, followed promptly by the onset of rapid quenching.  However, in order to elucidate the cause of this coalescence-era quenching, further work is required to measure properties such as the gas content, its spatial distribution and dynamical state.

\appendix

\section{Derivation of merger completion correction}

Following conventions in machine learning, we begin by defining the completeness ($C$) and purity ($P$) of our post-merger samples as a function of the true positives (TP: correctly labelled post-mergers), false positives (FP: non-mergers incorrectly labelled as post-mergers), true negatives (TN: correctly labelled non-mergers) and false negatives (FN: post-mergers incorrectly labelled as non-mergers):

\begin{equation}\label{c_eqn}
  C = \frac{TP}{TP+FN}
\end{equation}

and

\begin{equation}\label{p_eqn}
  P = \frac{TP}{TP+FP}.
\end{equation}

The denominators of of Equations \ref{c_eqn} and \ref{p_eqn} represent the number of intrinsic positives, $N_{i,p}$ (actual number of post-mergers) and the number of detected positives $N_{d,p}$ (number of galaxies classified as post-mergers).  Re-arranging and substituting terms gives

\begin{equation}\label{nip_eqn}
  N_{i,p} = \frac{P N{d,p}}{C}.
\end{equation}

The merger fraction that we measure ($f_{meas}$) is

\begin{equation}
f_{meas}=\frac{N_{d,p}}{N_T}
\end{equation}

where $N_T$ is the total number of galaxies in the sample.  What we want is the true merger fraction

\begin{equation}
  f_{true}=\frac{N_{i,p}}{N_T}.
\end{equation}

Substituting Equation \ref{nip_eqn} in the above yields

\begin{equation}
  f_{true}= \frac{P N_{d,p}}{C N_T}
\end{equation}

which can be more simply written as

\begin{equation}
  f_{true}=f_{meas} \frac{P}{C}.
\end{equation}

\end{document}